\documentclass[aps,twocolumn,prd,showpacs]{revtex4}

\usepackage{graphicx}

\usepackage{amsfonts}

\usepackage{amssymb}

\usepackage{array}
\usepackage{graphicx}
\usepackage{pstricks}
\usepackage{color}
\usepackage{xcolor}

\usepackage{subfigure}
\usepackage{amsmath}

\usepackage{latexsym}

\begin{document}

\title{Interference phenomenon and geometric phase for Dirac neutrino in $\pi^{+}$ decay}
\email[~Published~in: Phys.Rev.D {\bf 87}, 117302 (2013); \\
DOI: 10.1103/PhysRevD.87.117302]{} 
\author{J. Syska  }
\author{J. Dajka}
\author{J. {\L}uczka}
\affiliation{Institute of Physics, University of Silesia, 40-007
Katowice, Poland }

\begin{abstract}
We analyze the geometric phase in the neutrino oscillation phenomenon,
which follows the pion decay $\pi^{+} \rightarrow \mu^{+} + \nu_{\mu}$.
Its  value $\pi$ is consistent with the present-day global analysis of the Standard Model neutrino oscillation parameters,
accounting for the nonzero value of $\theta_{13}$. The impact of the charge-parity ($CP$) violating phase $\delta$, the neutrino's nature, and the new physics is discussed.
\end{abstract}

\maketitle

\vspace{-4mm}


\vspace{-4mm}

\label{Introduction}


{\bf I. Introduction.} The aim of this brief paper is to  discuss
the idea that in measurement subtleties of the neutrino oscillation  phenomenon, geometrical properties reflected in the geometric phase of  the  oscillating flavor neutrino are important.
In Ref.~\cite{Mehta} it was proposed that the production and detection of the neutrino shall be treated as the split-beam experiment in the energy space.
In the present paper, we consider  the muon neutrino which is produced in
the decay of pion to muon and the Dirac neutrino, namely $\pi^{+} \rightarrow \mu^{+} + \nu_{\mu}$ \cite{Giunti-Kim}.
The flavor neutrino state $|\nu_{\mu}\rangle$ is
a superposition of the stationary states  $|\nu_{i}\rangle_{\lambda} \equiv |p, \lambda, i\rangle$ of the definite masses $m_{i}$, $i=1,2,3$, helicities $\lambda=-1$ or $+1$, and four-momentum $p$.  When the new physics (NP) interactions
are included,
this superposition
composes
the mixed state \cite{OSZ,ZZS}. Thus, the flavor neutrino, here $\nu_{\mu}$,
represents
the beam of three massive states, which split at the moment of production of the $\alpha = \mu$-flavor superposition, propagate, and finally at the distance $L$, interfere in the detector in the  $\beta$-flavor interference pattern.
This interference experiment for the neutrino proposed in \cite{Mehta} and discussed in \cite{Mehta,DLS} in two flavor neutrino cases, allows us to test the dependence of the type \cite{sjuk2} of the Aharonov-Anandan  geometric phase (GP) \cite{A-A} on the particular field theory model to which this  paper is devoted.

The global analysis of neutrino oscillation parameters  \cite{global-fit} shows the discrepancy in the data for the atmospheric neutrino mixing angle $\theta_{23}$. For the normal neutrino mass
ordering (and we will use this one), the profile of the $\Delta \chi^2$ test statistics
has two almost equally deep minima--the
``local minimum'' ({\it lm}) for the solar plus reactor long-baseline
and accelerator long-baseline
neutrino experiments, with new data from the $\nu_{\mu}$ and $\bar{\nu}_{\mu}$ channels included,
and
the ``global minimum'' (\textsl{g}{\it m}), which includes  data from  atmospheric neutrinos, too.
The profile is practically symmetric
and the preference (if any, see \cite{Pascoli-Schwetz}) of \textsl{g}{\it m} (with $\sin^{2}\theta_{23} = 0.427$)
over {\it lm} (with $\sin^{2}\theta_{23} = 0.613$)
is very weak as the difference of $\Delta \chi^2$ in these minima is equal to $0.02$ \cite{global-fit}.
The 2$\sigma$   range  $(0.38, 0.66)$ covers both of them.  The experimental reason is that $\theta_{23}$ strongly depends on the $CP$ violating phase $\delta$ \cite{Pascoli-Schwetz}, whose 1$\sigma$ range is $\langle 0, 2 \pi)$ \cite{global-fit}.
For further discussion of this problem,  see \cite{theta23-next,theta23-atmospheric}.
It will appear that the mean $\sin^{2}\theta_{23} \approx 0.517$
is the robust one. The central values of the other oscillation parameters  are \cite{global-fit}
$\sin^{2}\theta_{12} = 0.320$, $\sin^{2}\theta_{13} = 0.0246$, $\delta = 0.80 \, \pi$, $\Delta m_{21}^{2}= 7.62 \times 10^{-5}$$eV^2$ and $\Delta m_{31}^{2}= 2.55 \times 10^{-3}$$eV^2$.\\

{\bf A. Muon neutrino density matrix}: From the $\pi^{+}  \rightarrow \mu^{+}  + \nu_{\mu}$ decay experiments \cite{Giunti-Kim}, we know that the fraction of the right-handed
$N_{\nu_{+1}}$ to the left-handed  $N_{\nu_{-1}}$ neutrinos fulfils the constraint  \cite{PDG_epsR,polar}
\begin{eqnarray}
\vspace{-2mm}
\label{N+1 do N-1}
N_{\nu_{+1}}/N_{\nu_{-1}} < 0.002 \; .
\vspace{-2mm}
\end{eqnarray}
Let us assume that the pion
decays effectively both
in the left ($L$) and right ($R$) chiral charge current ($CC$) interactions \cite{ZZS}   via the exchange of the
Standard Model ($\nu$SM) $W$ boson only.
Then, at the $W$-boson energy scale, the $R$ and $L$ chiral pion decay constants
\cite{Berman-Kinoshita}
are equal \cite{C-M}. Moreover,  the pseudoscalar correction to the pion hadronic matrix element can be neglected due to its smallness  \cite{EGPR}.  Then the invariant amplitudes $A_{i}^{\mu}{ ^{\; \lambda; \lambda_{\mu}}}(p)$ in the decay
$\pi^{+} \rightarrow \mu^{+} + \nu_{i,\lambda}$
are related as follows \cite{Giunti-Kim,ZZS},
\vspace{-2mm}
\begin{eqnarray}
\label{amplitudes in pion decay -+1}
|A_{i}^{\mu}{ ^{\; +1; +1}}(p)|^2 = |A_{i}^{\mu}{ ^{\; -1; -1}}(p)|^2 \, \frac{|\varepsilon_{R}|^{2} | U_{\mu i}^{R}|^{2} }{|\varepsilon_{L}|^{2} | U_{\mu i}^{L} |^{2} }\; .
\vspace{-3mm}
\end{eqnarray}
Here, $U_{\alpha i}^{L}$ and  $U_{\alpha i}^{R}$ are the $L$ and $R$ chiral neutrino mixing matrices, which enter into the CC Lagrangian
in the products with the coupling constants $\varepsilon_{L}$ and $\varepsilon_{R}$,
respectively  \cite{ZZS}. The NP values of $\varepsilon_{L}$ and $\varepsilon_{R}$ can deviate slightly from the $\nu$SM values 1 and 0,  respectively. However, the Fermi constant constraint $\varepsilon_{L}^4 +
\varepsilon_{R}^4 = 1$ should hold.

Under the above conditions, in the process of neutrino production (P)
the nonzero neutrino density matrix elements in the mass-helicity basis $ |\nu_{i} \rangle_{\lambda} $
and in the center-of-mass (CM) frame are as follows \cite{OSZ,ZZS}:
\vspace{-1mm}
\begin{eqnarray}
\label{production matrix - - + +}
\!\!\!\!\!\!\!\! \varrho^{{\rm P} \mu \; i; \, i'}_{-1; \,
-1} \! = \! \frac{|\varepsilon_{L}|^{2} U_{\mu i}^{L \ast} U_{\mu i'}^{L}}{ |\varepsilon_{R}|^{2} + |\varepsilon_{L}|^{2}}  , \;
\varrho^{{\rm P} \mu \; i; \, i'}_{+1; \, +1} \! = \! \frac{
|\varepsilon_{R}|^{2} U_{\mu i}^{R \ast} U_{\mu i'}^{R} }{ |\varepsilon_{R}|^{2} + |\varepsilon_{L}|^{2}}    \; ,
\vspace{-2mm}
\end{eqnarray}
constituting the muon neutrino $6 \times 6$-dimensional block diagonal density matrix $\rho^{{\rm P} \mu} = {\rm diag}(\varrho^{{\rm P} \mu}_{-1; \,
-1}, \varrho^{{\rm P} \mu}_{+1; \,
+1}) $ with two $3 \times 3$ matrices given in (\ref{production matrix - - + +}).
Here we choose $U_{\alpha i}^{R} = U_{\alpha i}^{L} = U_{\alpha i}$, where $U$ is the
Maki-Nakagawa-Sakata
neutrino mixing matrix \cite{MNSP}, as the full statistical analysis of this hypothesis is beyond the data accessible in the present-day experiments.
Using (\ref{N+1 do N-1})
we obtain the
bound on the ratio
$|\varepsilon_{R}/\varepsilon_{L}| < 0.0447$.
It constrains
the density matrix $\varrho^{{\rm P}  \mu \; i; \, i'}_{+1; +1}$ of the   initial neutrino. Its evolution and the effective Hamiltonian during the neutrino propagation are described in the next section.
\\
Next, for the neutrino energy $E_{\nu}>$ 100 MeV,  the neutrino is in practice the relativistic particle.  Hence  the effect of the helicity Wigner rotation is negligible \cite{OSZ} and the result for the density matrix in the laboratory ({\bf L})
frame is $\varrho_{\rm {\bf L}}^{\rm P}(\vec{p}_{\rm {\bf L}}) = \varrho^{\rm P}(\vec{p})$.
Finally, only the neutrino which is produced in the {\bf L} frame in the forward direction along the $z$ axis
reaches the detector and we choose this axis as the quantization one.  \\

{\bf II. Evolution of the density matrix.} Under the requirement of the nondissipative homogeneous medium,  the Liouville–-von Neumann equation governs the density matrix evolution. Thus, in the ultrarelativistic case, when the distance and the propagation time approach the relation $z=t$, the evolution rule for the neutrino density matrix is as follows:
\begin{eqnarray}
\label{evolution rho from T and P}  \!\!
\rho^{\mu}(t=0) \rightarrow \rho^{\mu}(t)  = e^{-i \, {\cal{H}} \,t  }  \rho^{{\rm P} \mu}(t=0) \; e^{i \, {\cal{H}} \,t  } \, ,
\end{eqnarray}
where $\rho^{{\rm P} \mu}$ is an initial density matrix
(\ref{production matrix - - + +}) and ${\cal{H}}$ is the effective Hamiltonian.

With three massive and two helicity neutrino
states, the effective Hamiltonian
${\cal{H}}$
has the
$6\times6$-dimensional representation. In the case of the axial-vector
interactions only, the effective Hamiltonian  ${\cal H}$
can be considered as block diagonal  with two $3 \times 3$ matrices,
\begin{eqnarray}
\label{Eff Ham}
{\cal{H}} =
\mathcal{M} + {\rm diag}({\cal{H}}_{--}, {\cal{H}}_{++}) \, .
\end{eqnarray}
Here $\mathcal{M} = {\rm diag}(E_{1}^{0}, E_{2}^{0},
E_{3}^{0},E_{1}^{0}, E_{2}^{0}, E_{3}^{0})$ with $E_{i}^{0} = E_{\nu} + m_{i}^{2}/ 2 E_{\nu}$ ($i =1,
2, 3$) is the mass term, where $E_{\nu}$ is the energy for the massless neutrino
\cite{Giunti-Kim}.
The  interaction Hamiltonians for the coherent Dirac neutrino scattering inside unpolarized matter read  \cite{AZS,Giunti-Kim,ZZS}
\begin{eqnarray}
\label{H0D}
({\cal{H}}_{- -})_{ij} &=& \sqrt{2}G_F
N_{e}  |\varepsilon_L|^2
\;U_{ei}^{L \ast}\; U_{ej}^{L} \,
\\
({\cal{H}}_{+ +})_{ij} &=& \sqrt{2}G_F \biggl\{ N_{e}  |\varepsilon_{R}|^2 \;U_{ei}^{R \ast}\;
U_{ej}^{R}
- \frac{\rho}{2}  N_{n} \, \varepsilon_{R}^{N \nu} \Omega_{ij}^{R} \biggr\}
\, , \nonumber
\end{eqnarray}
where $N_{e}$ and $N_{n}$
stand for the number of
background electrons ($e$) and neutrons ($n$) per unit volume, respectively and $\varrho
\simeq 1 $. Small NP deviations of the neutral coupling constant
for background particles are also neglected.
We choose the right chiral neutral mixing matrix in the mass basis
equal to  $\Omega^{R} = 3 \, {\rm diag}(w_{1}, w_{2}, w_{3})$ with $w_{1,2,3}=m_{e,\mu,\tau}/(m_{e}+m_{\mu}+m_{\tau})$, where  $m_{\alpha}$, $\alpha = e, \mu, \tau$, is the mass of electron, muon and tau lepton, respectively.
The bound on the neutrino right chiral neutral current ($NC$) coupling constant equal to
$|\varepsilon_{R}^{N \nu}| = 1$ can be obtained from the analysis of the charge-parity-time reversal ($CPT$) symmetry violation in the
neutrino oscillation survival events  \cite{Gonzalez_Maltoni_ogr_param}.
In the ana\-lysis we assume that the relevant $\nu$SM and NP coupling constants are real. \\
%





{\bf III. Analysis of geometric phase. }
Various types of geometric phases have been studied for a long time in  physical systems ranging from classical mechanics to high-energy physics \cite{darek}. There are also examples of exploiting the notion of  geometric phases in neutrino physics. Let us mention a few of them. In \cite{Berry in neutrino}, it was shown  that in the neutrino oscillations analysis, carried out under adiabatic conditions \cite{Giunti-Kim},  the nonzero Berry phase \cite{Berry} appears in the $\nu$SM  if a background consists of at least two varying densities.
The case of the three-level neutrino systems  was considered in \cite{Nakagawa}.
In \cite{Mehta} it was noted that the
Pancharatnam phase \cite{Pancharatnam}, which defines the relative phases between  states in the Hilbert space,
leads in two-flavor neutrino oscillation to the topological phase of the interference term, which is equal to zero or $\pi$ for the survival and appearance probability, respectively.

In the present case,  the neutrino $\nu_{\mu}$ is produced in the $\pi^{+}$ decay and propagates in the ordinary matter of the
crust (with the density $\rho=2.6$ ${\rm g/cm^3}$).  It reaches the detector after one oscillation period,  i.e. at the
maximum of the
survival transition rate $P(\nu_{\mu} \rightarrow \nu_{\mu})$.
If the detector lies  at the distance $L=800$ km which is
the baseline for the NO$\nu$A--Low-Z Calorimeter experiment \cite{Giunti-Kim}, it happens for $E_{\nu} = 0.803$ GeV
(what matters is the ratio $L/E_{\nu}$).
For the central value of the $CP$ violating phase $\delta = 0.80 \, \pi$ of the $U$-matrix, we obtain $P(\nu_{\mu} \rightarrow \nu_{\mu}) \approx 0.992$ (\textsl{g}{\it m}~) or $P(\nu_{\mu} \rightarrow \nu_{\mu}) \approx 0.991$ ({\it lm}~). For perfect cyclicity $P(\nu_{\mu} \rightarrow \nu_{\mu})=1$.
Hence the evolution is not exactly cyclic. Another measure of  the deviation from perfect cyclicity is the trace distance between initial state at $t=0$ and the state at time $t = L$  \cite{nielsen},
\begin{eqnarray}\label{trejs}
D=\frac{1}{2}||\rho^{\mu}(t=0)-\rho^{\mu}(t=L)|| \; ,
\end{eqnarray}
where the norm $||\varrho||=\mbox{Tr}\sqrt{\varrho^\dagger\varrho}$.
For  perfectly  cyclic evolution, $D=0$.
The calculations show that depending on $\delta$ and at the central values of  other parameters \cite{global-fit}, the trace distance $D \in  (0.012, 0.092)$ (\textsl{g}{\it m})
with the minimum  for $\delta=0$ and maximum for $\delta=\pi$
(the cases when  $CP$ is not violated). For $\delta=0.80$ the minimal value  $D=0.089$ (\textsl{g}{\it m}) is at $L=800$ km, which is the period of the oscillation.  The same is true for the ``local minimum''  \cite{global-fit,Pascoli-Schwetz}.
%
%
The deviation from the perfect cyclicity is due to the fact that the neutrino flavor state is a three-state system and
is not an eigenvector of the effective Hamiltonian governing its propagation.

In this paper, we exploit the  kinematic approach to the geometric phase \cite{sjuk2} which can be applied to arbitary (also nonunitary and/or  noncyclic) quantum evolution. It
possesses the following fundamental features~\cite{sjuk2}:  it is   gauge invariant, purification independent, and it reduces to well establish results in the limit of unitary evolution.  This approach has already been utilized in \cite{DLS}  for the two-flavor neutrino system both for nondissipative and dissipative cases.

In order to analyze the GP, it is convenient to present the density matrix  (4) in the spectral-decomposition form
\begin{eqnarray}
\label{spect}
\rho^{\mu}(t)=\sum_{i=1}^{6} \, \lambda_{i}^{\mu}(t) \, |w_{i}^{\mu}(t)\rangle \langle w_{i}^{\mu}(t)|,
\end{eqnarray}
where $\lambda_{i}^{\mu}(t)$ and $ |w_{i}^{\mu}(t)\rangle $ are the eigenvalues and eigenvectors of the matrix $\rho^{\mu}(t)$.
Then the geometric phase $\Phi^{\mu}(t)$ at time $t$ associated with such an evolution is defined by the following relation  \cite{sjuk2}:
\begin{eqnarray}
\label{Phi}
\Phi^{\mu}(t) &=& \mbox{Arg}\left[\sum_{d=1}^{6} [\lambda_d^{\mu}(0) \lambda_d^{\mu}(t)]^{1/2}\langle
w^{\mu}_d(0) |w^{\mu}_d(t)\rangle\right.\nonumber\\ && \left. \times
\exp(-\int_0^t \langle w^{\mu}_d(s)|\dot{w}^{\mu}_d(s)\rangle ds)\right],
\end{eqnarray}
where  $\mbox{Arg}\left[Z\right]$ denotes argument
of the complex number $Z$,  $\langle w^{\mu}_d|w^{\mu}_d \rangle$ is a scalar product, and the dot indicates the derivative with respect to time $s$. It is natural to  analyze the GP at time $t=L$, which corresponds to the
period of neutrino oscillations. Below, we study the GP at this time and  use the notation $\Phi \equiv \Phi^{\mu}(t = L)$.

In \cite{Mehta} it was assumed that
neutrino oscillation realizes a kind of interference experiment, and under this assumption it was proven that in the two-flavor case,
the topological phase of the interference term
is reflected in the orthogonality of the mixing matrix.
In the present paper, it is suggested that because this interference experiment
reflects the orthogonality of
the
neutrino mixing matrix, the GP takes the topological value $\pi$ (the correction from the $CP$ violating phase $\delta$ will appear very tiny). This value of GP influences self-consistently the parameters of the mixing matrix.\\

\begin{figure}[top]
\vspace{-2mm}
\begin{center}
\includegraphics*[width=.45\textwidth,height=53mm,
angle=0]{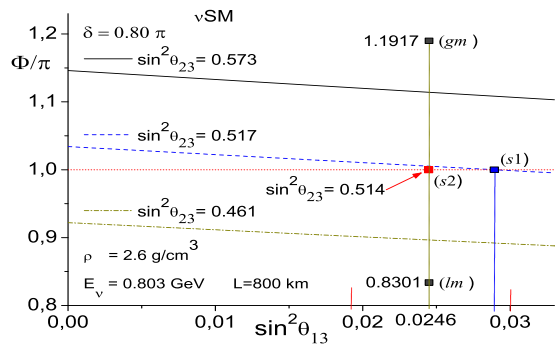}
\end{center}
\vspace{-10mm}
\caption{The geometric phase $\Phi \equiv \Phi^{\mu}(t = L)$
vs
$\sin^{2}\theta_{13}$ plotted for the value of  $\sin^{2}\theta_{23}=0.461$ ($+1\sigma$ bound {\it lm}~) and  $\sin^{2}\theta_{23}=0.573$ ($-1\sigma$ bound \textsl{g}{\it m}~)  \cite{global-fit}. For any
$\sin^{2}\theta_{13}$ the whole area on the figure is covered by the $2 \sigma$ range for $\sin^{2}\theta_{23}$.  The GP, 1.1917, and 0.8301, for the central values of \textsl{g}{\it m} and {\it lm}, respectively, are signified. Two examples of the GP solution are pointed out, $\sin^{2}\theta_{23}=0.517$ ({\it s1}) and $\sin^{2}\theta_{23}=0.514$ ({\it s2}).
The $\pm 2 \sigma$ limits, 0.019 and 0.030, for $\sin^{2}\theta_{13}$ are signified by the short vertical lines.
}
\label{fig-GP}
\vspace{-3mm}
\end{figure}

\begin{figure}[here]
\vspace{-3mm}
\begin{center}
\includegraphics*[width=.45\textwidth,height=53mm,angle=0]
{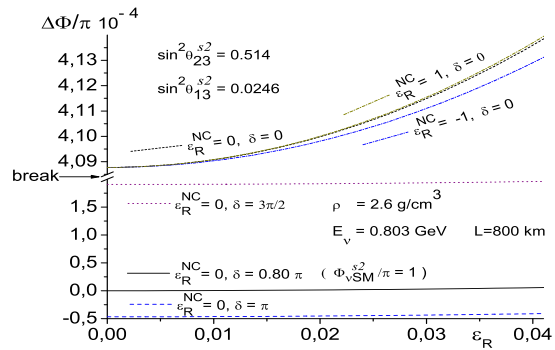}
\end{center}
\vspace{-10mm}
\caption{The geometric phase difference
 $\Delta \Phi = \Phi_{NP} - \Phi_{\nu {\rm SM}}$
as a function of  the
NP coupling constant  $\varepsilon_{R}$.
Each curve corresponds to one
$\delta$ and one NP
value of $\varepsilon_{R}^{N \nu}$.
As the reference level the $\nu$SM {\it s2} solution (see Fig.~1)
is taken
for which $\Phi_{\nu SM} = \pi$ (up to the influence of
$\delta=0.80 \pi$ equal to $-4.09 \times 10^{-4} \; \pi$). }
\label{fig-GP-dif}
\vspace{-2mm}
\end{figure}

{\bf A. Geometric phase in $\nu$SM}:
Because of the mentioned discrepancy in the data,  the analysis of the  GP given by Eq.(\ref{Phi}) is for $\nu$SM performed for {\it lm} and \textsl{g}{\it m} \cite{global-fit}.
The results are presented in Fig.~\ref{fig-GP}. The GP for the central values of {\it lm} and \textsl{g}{\it m} are equal to $\Phi^{\it lm}=1.1917 \, \pi$ and $\Phi^{\textsl{g}{\it}m}=0.8301 \, \pi$, respectively.
The bottom line is plotted for $\sin^{2}\theta_{23}=0.461$ for $+1\sigma$ bound of {\it lm} range $(0.400,0.461)$ and the upper one for $\sin^{2}\theta_{23}=0.573$ for $-1\sigma$ bound of \textsl{g}{\it m} range $(0.573,0.635)$ \cite{global-fit}. We notice that (with other oscillation parameters fixed) $\Phi$ changes linearly as the function of  $\sin^{2}\theta_{13}$, where $\theta_{13}$ is the third mixing angle of $U$ \cite{Giunti-Kim}.
Two examples of the  GP solution with $\Phi=\pi$ are pointed out, the first one for $\sin^{2}\theta_{23}=0.517$ ({\it s1}) and the second one  for $\sin^{2}\theta_{23}=0.514$ ({\it s2}).
The former value, $\sin^{2}\theta_{23}=0.517$, is the arithmetic mean of the $+1\sigma$ bound 0.461 for {\it lm} and $-1\sigma$ bound 0.573 for \textsl{g}{\it m}.
%
%
With this value, the condition of the geometric value $\Phi=\pi$ for the  GP gives $\sin^{2}\theta_{13} \approx 0.029$, which lies in the $2\sigma$ range  (0.019, 0.030) for $\sin^{2}\theta_{13}$ \cite{global-fit}.
In the second example the current central value $\sin^{2}\theta_{13} = 0.0246$ is chosen.  Now, the ``GP solution''  for $\Phi=\pi$ is $\sin^{2}\theta_{23}=0.514$ ({\it s2}) \cite{theta23-atmospheric}.
The value $\pi$ of the  GP arises as the result of the interference of the neutrino mass states \cite{Mehta} at the point of the flavor neutrino detection at the first period.
\\
The  GP changes mainly with  $\sin^{2}\theta_{23}$ and $\sin^{2}\theta_{13}$ (see Fig.~\ref{fig-GP}), whereas the impact of the other $\nu$SM oscillation parameters is significantly weaker. That is, the change of  $\Delta^{2}m_{31} \equiv m^{2}_{3} - m^{2}_{1}$, $\Delta^{2}m_{21} \equiv m^{2}_{2} - m^{2}_{1}$ and $\sin^{2}\theta_{12}$ in their $2\sigma$ ranges \cite{global-fit} causes the change of $\Phi/\pi$ approximately equal to $3 \times 10^{-4}$, $10^{-5}$, and $3 \times 10^{-6}$, respectively.
The small dependence
on the $CP$-violating phase $\delta$ is presented in Fig.~\ref{fig-GP-dif}. Its impact is of the order of $10^{-4}$.
The numerical calculations show that in
$\nu$SM with the period $L=800$ km the
GP takes the topological values
$\Phi^{\mu} = n \; \pi$
$({\rm mod} \,2 \, \pi)$,
$n \in N$ (up to the influence of the phase $\delta$).

Interestingly, for the old $\nu$SM global analysis  \cite{global-fit-old} with the central value $\sin^{2}\theta_{13} = 0.010$ (but when the non-zero value of $\theta_{13}$ was still disputed), the  GP analysis had suggested that the condition $\Phi=\pi$ requires $\sin^{2}\theta_{13}$ to be enlarged approximately to 0.0175, which value was then inside $\pm 1\sigma$ limits, or alternatively that $\sin^{2}\theta_{23}$ shall be diminished from 0.51  to 0.506 \cite{global-fit-old}.\\

{\bf B. Geometric phase in  NP}:
 The bounds on  the $CC$ and $NC$ right-chiral coupling constants $\varepsilon_{R}$ and  $\varepsilon_{R}^{N \nu}$  are given in the Introduction. In Fig. \ref{fig-GP-dif},
the difference $\Delta \Phi = \Phi_{NP} - \Phi_{\nu {\rm SM}}$ between NP and $\nu$SM values of $\Phi$ as the function of  $\varepsilon_{R}$
is depicted. Each curve corresponds to the different value of the
phase $\delta$. The upper impact of $\varepsilon_{R}$
on $\Phi/\pi$
is of $10^{-6}$ order.  Even weaker is the influence
of $\varepsilon_{R}^{N \nu}$. Yet, because it enters linearly into the Hamiltonian (\ref{H0D}) \cite{AZS}, it therefore depends
on the $\varepsilon_{R}^{N \nu}$ sign,  too.

\begin{figure}[here]
\vspace{-2mm}
\begin{center}
\includegraphics*[width=.45\textwidth,height=53mm,angle=0]
{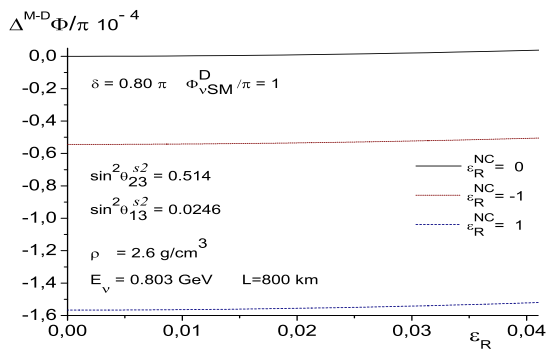}
\end{center}
\vspace{-8mm}
\caption{Comparison of geometric phases for the Majorana and Dirac NP neutrinos. The geometric phase difference
 $\Delta^{M-D} \Phi = \Phi^{M} - \Phi^{D}$
is depicted as a function of the
NP coupling constant  $\varepsilon_{R}$.
Each curve corresponds to one value of $\varepsilon_{R}^{N \nu}$.  }
\label{fig-GP-dif M-D}
\vspace{-3mm}
\end{figure}

Finally, let us comment on the Majorana neutrino case.
In the case of  $\nu$SM,  there is no difference between GP for the Dirac and Majorna neutrinos \cite{DLS}. In the case of NP,
 the difference $\Delta^{M-D} \Phi = \Phi^{M} - \Phi^{D}$ of the geometric phases $\Phi^{M}$ and $\Phi^{D}$ for the Majorana  neutrino and Dirac  neutrino is depicted in Fig. \ref{fig-GP-dif M-D} as a function of the NP couplings \cite{AZS}.  The impact of $\varepsilon_{R}^{N \nu}$ and $\varepsilon_{R}$
on $\Phi/\pi$ is of order $10^{-4}$  and $10^{-5}$--$10^{-6}$, respectively.\\

{\bf IV. Conclusions. } With this brief paper we have shown that selected  properties of
the nonadiabatic noncyclic flavor neutrino oscillation can be analyzed in terms of the type of the Aharonov-Anandan GP introduced in \cite{sjuk2}.
At first, using the trace distance $D$, it has been checked that
in one oscillation period, the muon neutrino state performs the evolution along the path in its
Hilbert space, which shows some small departure from cyclicity. Hence
the solid angle encircled in this space is close to $2 \, \pi$
(similar to the spin particle moving in the mesoscopic ring \cite{ring}).
This motivates the use of the kinematic approach to the geometric phase presented in \cite{sjuk2} which attaches the geometric phase to the Pancharatnam relative one.
%
%
%
As mentioned above,
the described pattern of the interference in the energy space of the massive neutrino states is highly possible \cite{Mehta,DLS}. This in \cite{Mehta} enables us to use the Pancharatnam relative phase for the explanation of the orthogonality of the two-flavor mixing matrix.
In \cite{DLS}, the behavior of the GP attached to it was analyzed. \\
In this paper
it is pointed out that the present-day global analysis
of the oscillation parameters  \cite{global-fit} is consistent with the GP value equal to $\pi$, which is the reflection of both the unitarity of the mixing matrix and the values of its experimentally estimated parameters.
%
%
The  GP is  sensitive to changes of $\sin^{2}\theta_{23}$ and $\sin^{2}\theta_{13}$ (see Fig.~\ref{fig-GP}), currently the more disputed parameters \cite{global-fit}, whereas the influence of the other $\nu$SM oscillation parameters is approximately of the relative order $10^{-6}$--$10^{-4}$ in their $2\sigma$ ranges.
The NP corrections
connected with the right-chiral $CC$ and $NC$ currents are at most
of the relative order of
$10^{-6}$, being at present far beyond the experimental  verification.
%
%
%
%
Recent progress in entirely novel experimental techniques makes the verification of presented findings more realistic in the future. In the long term, our research may provide new tools for analysis of neutrino physics.


{\bf Acknowledgments}: The work supported by the NCN Grants No. N202
052940 and No. DEC-2011/01/B/ST6/07197.


\end{document}